\documentclass{article}

\usepackage{PRIMEarxiv}

\usepackage[utf8]{inputenc} 
\usepackage[T1]{fontenc}    
\usepackage{hyperref}       
\usepackage{url}            
\usepackage{booktabs}       
\usepackage{amsfonts}       
\usepackage{nicefrac}       
\usepackage{microtype}      
\usepackage{lipsum}
\usepackage{fancyhdr}       
\usepackage{graphicx}       
\graphicspath{{media/}}     

\usepackage[numbers]{natbib}

\usepackage{graphicx} 
\usepackage{import}
\usepackage[table]{xcolor}
\usepackage{soul}
\usepackage{placeins}
\usepackage{float}
\usepackage{adjustbox}
\usepackage{multirow}
\usepackage{caption}
\usepackage{footnote}
\usepackage{threeparttable}
\floatstyle{plain}
\usepackage{listings}
\newfloat{listing}{htbp}{lop}

\usepackage[scaled=0.85]{beramono}
\usepackage[framemethod=TikZ]{mdframed}  
\usepackage{booktabs}   
\usepackage{amssymb}    
\usepackage{pifont}     
\usepackage{afterpage}

\newcommand{\yes}{\textcolor{teal}{\ding{51}}} 
\newcommand{\no}{\textcolor{lightgray}{\scriptsize$\cdot$}} 

\definecolor{rowgray}{gray}{0.95}

\definecolor{codegreen}{rgb}{0,0.6,0}
\definecolor{codegray}{rgb}{0.5,0.5,0.5}
\definecolor{codepurple}{rgb}{0.58,0,0.82}
\definecolor{backcolour}{rgb}{0.98,0.98,0.98}

\lstdefinestyle{mystyle}{
    commentstyle=\color{codegreen},
    keywordstyle=\color{magenta},
    numberstyle=\tiny\color{codegray},
    stringstyle=\color{codepurple},
    basicstyle=\ttfamily\footnotesize,
    breakatwhitespace=false,         
    breaklines=true,                 
    keepspaces=true,                 
    numbers=left,                    
    numbersep=10pt,                  
    showspaces=false,                
    showstringspaces=false,
    showtabs=false,                  
    tabsize=2,
    xleftmargin=20pt,  
    captionpos=b,              
    abovecaptionskip=10pt,     
}
 
\mdfdefinestyle{fancyframe}{
    backgroundcolor=backcolour,
    linecolor=black!20,
    linewidth=0.5pt,
    roundcorner=6pt,
    innertopmargin=5pt,
    innerbottommargin=5pt,
    innerleftmargin=10pt,
    innerrightmargin=5pt,
    skipabove=10pt,
    skipbelow=10pt,
    frametitlefont=\ttfamily\bfseries\small, 
    frametitlebackgroundcolor=black!5,       
    frametitlerule=true,                     
    frametitlerulewidth=0.5pt,
    frametitlerulecolor=black!20
}

\newcommand{\checksym}[1]{\textcolor{teal}{$\checkmark$}}
\newcommand{\timessym}[1]{\textcolor{red}{$\times$}}
\newcommand{\projectname}[1]{SMDT}

\title{Social Media Data Toolkit: Standardization and Anonymization of Social Network Datasets
}

\author{
  Ali Najafi \\
  Sabanci University \\
  Türkiye\\
   \And
  Letizia Iannucci\\
  Aalto University \\
  Finland\\
  \And
  Mikko Kivelä\\
  Aalto University \\
  Finland\\
  \And
  Onur Varol\\
  Sabanci University \\
  Türkiye\\
}

\begin{document}
\maketitle
\begin{abstract}
The rapid diversification of social media platforms and the increasing restrictions on official APIs have significantly complicated cross-platform analysis. Researchers are often forced to rely on heterogeneous datasets obtained through web scraping and historical archives; however they often lack structural consistency. Prior to conducting cross-platform social media analyses, one needs to answer three critical questions: (1) What makes platforms different and similar? (2) How were the datasets collected? (3) How can we align the datasets of different platforms to conduct fair analyses? To address these questions, we introduce the Social Media Data Toolkit (\projectname{}), a comprehensive Python framework designed for the standardization, anonymization, and enrichment of social network datasets. \projectname{}  unifies diverse data structures into a generic schema comprising Communities, Accounts, Posts, Actions, and Entities to facilitate multi-platform research. The framework features a configurable anonymization module to secure Personally Identifiable Information (PII) and an extendable enrichment layer that integrates Large Language Models (LLMs) and network analysis tools for downstream tasks such as stance detection and toxicity scoring without creating codebase for different datasets. We demonstrate the versatility of  \projectname{}  through four case studies spanning from textual analysis of the content to network analysis across platforms.
To offer reproducible social media research, \projectname{} is released as an open-source tool featuring detailed documentation and practical guides for researchers at any skill-level. It can be accessed at \href{https://github.com/ViralLab/SMDT}{github.com/ViralLab/SMDT} and \href{https://varollab.com/SMDT}{varollab.com/SMDT}.
\end{abstract}


\section{Introduction}

The Internet and social media platforms have become a central part of our lives. As of April 2024, there are 5.44 billion internet users worldwide and 5.07 billion or 62.6 percent of the world's population, were on at least one social media platform according to a recent report by Statista.\footnote{https://www.statista.com/statistics/617136/digital-population-worldwide}
Considering the wide range of adoption, it is not a surprise to see people turning social media to learn about news \citep{hermida2012share,kumpel2015news,newman2011mainstream}, seek for an advice on health \citep{chen2021social,cole2016health} and finance \citep{kedvarin2023social,renault2017market}, and interact with other accounts to follow their updates and share content with their own network.
European Union making efforts to ensure safety of these online platforms, they try to monitor and regulate for online manipulation and coordinated activities. Very large online platforms (VLOPs) and search engines with  more than 45 million users are especially an interest and EU is developing programs like Digital Services Act (DSA) to investigate social media and also create channels for researchers to conduct research with these datasets. 

However, the landscape of computational social science is shifting drastically. We have entered a ``Post-API'' era, where the stable, official access points that researchers once relied on are disappearing \citep{freelon2018computational}. While major platforms historically offered documented APIs for academic research, the ``APIcalypse'' has seen companies like X (formerly Twitter) and Facebook restrict or monetize access, effectively privatizing public data \citep{bruns2021after}. This shift poses a severe risk to scientific equity, as high-quality data becomes a luxury available only to well-funded institutions, while others are left to navigate a fragmented landscape of ``fragile'' data \citep{abbott2024applications}.

Facing such constraints, researchers must now navigate a highly fragmented ecosystem of data acquisition \citep{freelon2018computational,poudel2024navigating}. Rather than relying on a single, standardized access point, datasets are increasingly cobbled together from a multitude of sources and collection methods. These include remnant or commercial APIs, web scraping, official data dumps (often mandated by regulations like the DSA), and voluntary data donations from users. Each of these methods introduces its own structural quirks, limitations, and legal considerations. For instance, while scraping is highly sensitive to user interface changes and yields structurally inconsistent data, official data dumps may use entirely different, platform-specific schemas, and donated data often requires rigorous anonymization. Consequently, the primary hurdle for modern computational social science is reconciling this sheer heterogeneity of data formats and sources into a cohesive structure suitable for analysis, sharing, and long-term reproducibility \citep{Katharina2017Archiving}.

On the other hand, we see emergence of new social network platforms over the years \citep{seckin2025rise,mekacher2024koo,gerard2023truth}. Some platforms acquire large user-bases, while others disappear over time. Competition of the platforms are fierce and some platform try to offer different functionalities to keep their user bases engaged, while other thrive by offering a platform for certain fringe communities \citep{shah2024unfiltered,baele2021variations}. 
Some platforms prevent their users from posting on the platforms or suspend their accounts. 
These form of online censorship called deplatforming when used on a group of individuals or popular accounts \citep{rogers2020deplatforming,jhaver2021evaluating}.

We see examples of deplatforming on popular platforms like Reddit and Twitter. One of the most memorable example of deplatforming happened on Twitter when the platform bans former US president Donald J. Trump \citep{bryanov2021other,buntain2023cross}. 
Reddit also used deplatforming in 2015 and banned several communities for violating the site's anti-harassment policy \citep{chandrasekharan2017you}. 
These events led to migration of users from one platform to another; however, multi-platform analyses are challenging since accessing and processing datasets require different systems tailored for data formats provided their APIs or available information for scraping.

To conduct social media analyses, there are open-source software such as OSoME \citep{aiello2016osome} developed by Indiana University Network Science Institute. OSoME provides researchers to analyze social data. This software is mainly designed for the Twitter Streams and gives the users the flexibility to conduct temporal analyses, network analyses with visualization and geographic maps visualizations. As mentioned, this tool was mainly designed for one specific platform - Twitter. Other platforms like Communalytic by Social Media Lab at Toronto Metropolitan University is commercially available or offer academic license to provide no-code data collection services and tools for various platforms \citep{gruzd2020studying}.

Considering the rapidly shifting landscapes of social media, there is an emerging need to conduct cross-platform studies by merging datasets by shared topics or communities. However, this is a great challenge especially when data collection was not intended to conduct particular research study. Researchers either have access to existing datasets or they develop data collection pipelines for certain events. For instance, projects for gathering multi-platform datasets are available for US and Italy \citep{aiyappa2023multi,pierri2023ita}. Researchers are also conducting cross-platform studies to investigate ideological fragmentation \citep{di2025ideological} and engagement with partisan and low-quality news \citep{mosleh2025divergent}.

Another challenge to conduct analysis on multi-platform datasets is their different structure and mechanisms. A recent project, called MADOC,\footnote{\href{https://github.com/atomashevic/pyMADOC}{https://github.com/atomashevic/pyMADOC}} provides datasets for Twitter, BlueSky, Voat, and Koo in a standardized format \citep{dankulov2025multi}. Unfortunately, this tools offers only an interface to access existing standardized data but not a library to transform existing data.

Prior to conducting cross-platform social media analyses, one needs to answer three critical questions: (1) What makes platforms different and similar? (2) How were the datasets collected? (3) How can we align datasets of different platforms to conduct comparable analyses? Although definitions of concepts and functionalities may differ between platforms, each consist of core abstracted units such as accounts and content. Mechanisms to govern interactions are also rooted from same principles and can be standardized. More importantly, problems observed on these platforms such as misinformation, automated activities, and coordinated campaigns can also be observed across platforms and use similar strategies. 
In this work, we make the following contributions by developing the Social Media Data Toolkit framework:

\begin{itemize}
    \item \textbf{Data standardization.} We developed a framework that process datasets obtained from different platforms and transform them into one unified format. 

    \item \textbf{Anonymization of sensitive information.} Distribution of social media data requires anonymization of sensitive and personally identifiable information. Our framework process user-defined entities and replace them with hashed content.
    
    \item \textbf{Extendable modules.} Our code-base offers developers an ability to introduce their own models for analyzing account or content level information. Also, extending the analysis to new platforms is quite trivial thanks to the abstractions provided by \projectname{}. 

    \item \textbf{Documentation and recipes.} We offer reproducible analysis and recipes to share with other researchers on our website: \href{https://varollab.com/SMDT}{varollab.com/SMDT}.
\end{itemize}

\section{Social Media Platforms And Data}
The growing number of social media platforms presents significant challenges for developing cross-platform systems, as each service offers unique features and user dynamics. This work aims to develop a unified framework capable of incorporating various datasets and transforming them into a standardized format. To achieve this, as mentioned previously, we need to address the three critical questions which can help us in designing such a system and gives the researchers the idea of how to use such system.

\textbf{Question 1: What makes platforms different and similar?} 
Social media architectures vary significantly, enforcing distinct interaction patterns. For instance, threaded discussion platforms like Reddit require users to submit content within specific communities (subreddits), utilizing nested comments and scoring systems to organize discourse. In contrast, microblogging platforms such as Twitter (X) and Truth Social prioritize user-centric feeds; they distribute content to followers without rigid categorization, though topics are often aggregated via hashtags. Finally, Instant messaging platforms like Telegram and WhatsApp, although they are primarily designed for direct communication, increasingly facilitate content distribution through public channels and groups where creators broadcast to subscribers. 

\textbf{Question 2: How were the datasets collected?}
Another critical dimension in building a unified architecture is the methodology of data collection, which generally follows two paths: API Access and Web Scraping. Historically, platforms were more generous, providing API access to allow developers to build applications and researchers to study user behavior. More recently, for platforms that lack official API access, researchers must rely on web scraping to collect information from these services.

Another dimension that we should consider is whether the collected data was a historical data or collected via stream. Data acquisition characteristics also vary significantly depending on the provider. Platforms such as Bluesky offer real-time stream data via ``Firehose'' APIs. While this data is immediate, it often lacks comprehensive metadata regarding users or the full context of the content. Conversely, historical datasets provide a retrospective view that typically exclude any content deleted prior to the moment of collection and provide metadata reflecting at the time data collection.

Data richness is another critical factor when conducting these analyses. In their survey, \citet{guo2024survey} curated available datasets for information diffusion tasks. Some datasets are gathered primarily for content-level analysis and lack account-level information, whereas others are compiled specifically for network-level analysis.

\textbf{Question 3: How can we align the datasets of different platforms to conduct comparable analyses?} 
To answer this core question, we must identify common abstractions for the diverse elements present across different platforms and represent the data within those shared abstractions. Sometimes this alignment is trivial; for instance, a functionality might be called a "retweet" on one platform and a "ReTruth" on another, but both are fundamentally just platform-specific terms for the universal action of "reposting." In other scenarios, the mapping is significantly more complex, such as attempting to align the nuanced concept of "communities" across platforms that have vastly different underlying interaction mechanics. Furthermore, we must account for the structural inconsistencies and shortcomings stemming from diverse data collection methods. To address these standardization challenges, we introduce \projectname{}, a framework specifically designed to robustly map heterogeneous platform data into these unified, common abstractions.

\section{\projectname{} for multi-platform analysis}
\begin{figure}[t]
    \centering
    \includegraphics[width=\linewidth, keepaspectratio]{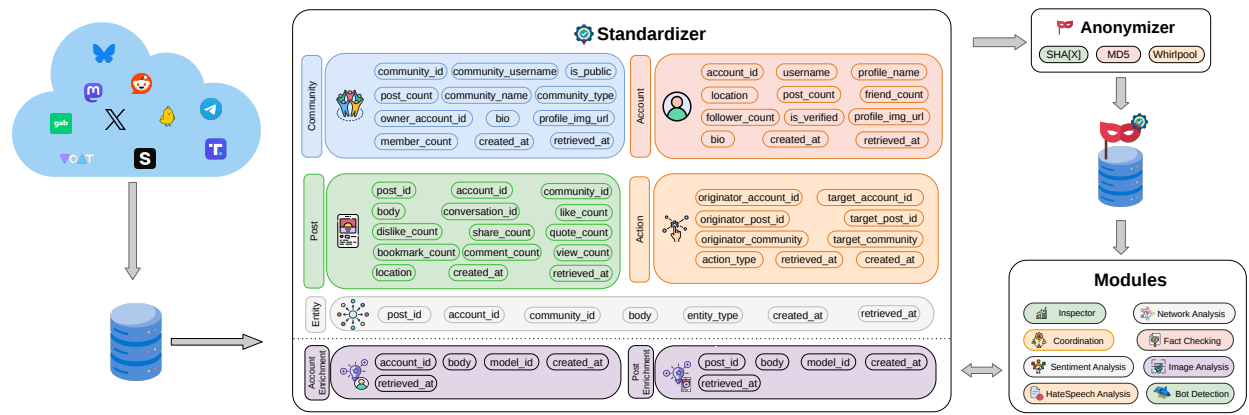}
    \caption{\textbf{Schematic of the \projectname{} platform}. Platform consist of distinct modules offering different capabilities. Standardizer unifies data coming from diverse social media platforms into one shared taxonomy. To distribute data for additional analysis, we anonymize the sensitive entries and store it further analysis. We offer different extension to enrich dataset by inferring account- and content-level measures such as bot detection and manipulated image detection. These base information and enrichments are then used for various down-stream analysis tasks such as coordination detection.}
    \label{fig:schema}
\end{figure}

\projectname{} consist of multiple modules which conforms with the software engineering best practices to provide a flexible, extendable, and scalable framework for conducting social media analyses. Figure~\ref{fig:schema} shows the high level structure of our framework. \projectname{} consists the following components: Standardizer, Inspector, Anonymizer, and Enrichers. In the following sections we will describe the functions of each component in details.

\subsection{Data standardization}

Social media datasets exhibit significant structural heterogeneity. While some platforms provide APIs that segregate objects—such as posts, users, and networks—others stream activities combined with entity metadata. Furthermore, web scraping efforts often result in isolated tabular structures.

The data standardization phase addresses this structural diversity by mapping disparate formats into a unified schema optimized for downstream analysis. Our objective is to populate the fields defined in Table \ref{tbl:basicforms} and store the processed data in a PostgreSQL database.

To facilitate access, our framework provides a robust database interface for interacting with these normalized datasets, including support for JSON exports to ensure interoperability with external toolsets. The proposed generic schema consists of five core tables: \texttt{Communities}, \texttt{Accounts}, \texttt{Posts}, \texttt{Actions}, and \texttt{Entities}, as well as two auxiliary tables: \texttt{AccountEnrichments} and \texttt{PostEnrichments}. This structure is specifically designed to support both single-platform and cross-platform analysis.

\begin{table}[h!]
\fontsize{8pt}{8pt}\selectfont
\centering
\caption{\textbf{Standardized format of data and its features.} There are five main categories for social media data to capture main interaction entities and their actions on social media platforms.}
\label{tbl:basicforms}
\begin{threeparttable}
\begin{tabular}{|c|l|}
  \hline
  \multicolumn{2}{|c|}{\textbf{Communities}} \\
  \hline
  \texttt{community\_id} & Unique identifier of a community \\
  \texttt{community\_type} & Channel or Group \\
  \texttt{community\_username$\dagger$} & Username of the community \\
  \texttt{community\_name$\dagger$} & Title of the community \\
  \texttt{bio$\dagger$} & Description of the community \\
  \texttt{is\_public$\dagger$} & Whether the community is public or not \\
  \texttt{member\_count$\dagger$} & Number of accounts in the community \\
  \texttt{post\_count$\dagger$} & Number of posts in the community \\
  \texttt{profile\_image\_url$^\dagger$} & Profile Image URL \\
  \texttt{owner\_account\_id$\dagger$} & Owner Account ID \\
  \texttt{created\_at} & Timestamp for the community creation date \\
  \texttt{retrieved\_at} & Timestamp for the community collection date \\ 
  \hline
  \multicolumn{2}{|c|}{\textbf{Accounts}} \\
  \hline
  \texttt{account\_id} & Unique identifier of an account. \\
  \texttt{user\_name$^\dagger$} & Unique profile name selected by the user \\
  \texttt{profile\_name$^\dagger$} & Name selected by account to visualize in the profile \\
  \texttt{bio$^\dagger$} & Description  \\
  \texttt{location$^\dagger$} & Geo Coordinates  \\
  \texttt{post\_count$^\dagger$} & Number of posts linked to account \\
  \texttt{friend\_count$^\dagger$} & Number of friends account have \\
  \texttt{follower\_count$^\dagger$} & Number of followers account have \\
  \texttt{profile\_image\_url$^\dagger$} & Profile Image URL \\
  \texttt{created\_at} & Timestamp for the account creation date \\
  \texttt{retrieved\_at} & Timestamp for the account collection date \\
  \hline
  \multicolumn{2}{|c|}{\textbf{Posts}} \\
  \hline
  \texttt{post\_id} & Unique identifier of the retrieved post. \\
  \texttt{account\_id} & Unique identifier of an account. \\
  \texttt{conversation\_id} & Thread’s originating post ID. \\
  \texttt{community\_id} & Community identifier of the post\\
  \texttt{body} & Information conveyed in the post as a text form \\
  \texttt{location$^\dagger$} & Geo Coordinates  \\
  \texttt{like\_count$^\dagger$} & Number of likes \\
  \texttt{dislike\_count$^\dagger$} & Number of dislikes \\
  \texttt{view\_count$^\dagger$} & Number of views \\
  \texttt{share\_count$^\dagger$} & Number of shares \\
  \texttt{comment\_count$^\dagger$} & Number of replies \\
  \texttt{quote\_count$^\dagger$} & Number of quotes \\
  \texttt{bookmark\_count$^\dagger$} & Number of bookmarks \\
  \texttt{created\_at} & Timestamp for the post creation date \\
  \texttt{retrieved\_at} & Timestamp for the post collection date \\
  \hline
  \multicolumn{2}{|c|}{\textbf{Entities}} \\
  \hline
  \texttt{post\_id} & ID of a post that contains an entity \\
  \texttt{body} & Information conveyed in a post \\
  \texttt{entity\_type} & Type of Entity (Hashtag, Link, etc.) \\
  \texttt{created\_at} & Timestamp for the interaction \\
  \texttt{retrieved\_at} & Timestamp for the post collection date \\
  \hline
  \multicolumn{2}{|c|}{\textbf{Actions}} \\
  \hline
  \texttt{originator\_account\_id} & ID of an account that initiates interaction \\
  \texttt{originator\_post\_id} & ID of a post that initiates interaction \\
  \texttt{target\_account\_id} & ID of an account that was effected \\
  \texttt{target\_post\_id} & ID of a post that was effected \\
  \texttt{action\_type} & Type of interaction like mention, repost, follow, etc. \\
  \texttt{created\_at} & Timestamp for the interaction \\
  \texttt{retrieved\_at} & Timestamp for the action collection date \\
  \hline\hline
  \multicolumn{2}{|c|}{\textbf{AccountEnrichments/PostEnrichments}} \\
  \hline
  \texttt{account\_id / post\_id} & ID of the account or post \\
  \texttt{model\_id} & Object types such as account or post \\
  \texttt{body} & JSON formatted responses of the inference models \\
  \texttt{created\_at} & Timestamp for the enrichment creation date \\
  \texttt{retrieved\_at} & Timestamp for the enrichment retrieval date\\
  \hline
\end{tabular}

\vspace{1mm}
\begin{tablenotes}
       \item [$\dagger$] Time dependent features captures values at particular timestamp.
       \item [$\ddagger$] Static models may increase versions incrementally and dynamics models may link it with timestamp of the model creation.
     \end{tablenotes}
\end{threeparttable}
\end{table}

\vspace{1em}

\noindent \textbf{\texttt{Communities}} table establishes a foundational taxonomy for digital environments, categorizing spaces according to their underlying communication topologies. We set out a primary distinction between Channels—broadcast-oriented, one-to-many architectures optimized for unidirectional information dissemination (\textit{e.g.}, Telegram Channels)—and Groups, which act as mesh networks fostering dense, many-to-many discursive interactions. Furthermore, this schema supports hierarchical nesting: Groups may function as autonomous conversation roots (\textit{e.g.}, Reddit submissions, Twitter/X threads) or exist as dependent sub-components embedded within the broader scope of parent Channels.  

\vspace{1em}

\noindent \textbf{\texttt{Accounts}} table stores user profile metadata functioning as a temporal snapshot, preserving the state of the user profile at the precise moment of data collection. Consequently, there may be a discrepancy between the reported \texttt{post\_count} attribute in the user’s profile and the actual number of records available in the \texttt{Posts} table, as the latter depends on data sampling and API retrieval limits.

\vspace{1em}

\noindent \textbf{\texttt{Posts}} table captures the content generate by the users listed in Accounts table and functions as the primary persistent store for the content layer of the ecosystem. It keeps the core message body alongside its associated metadata attributes.  This schema implements a unified data model to handle heterogeneous content sources, treating distinct entity types—such as a Twitter (X) status update or a Reddit platform comment—as structurally equivalent records. A critical architectural detail is the temporal dependency of the metadata: attributes like \texttt{like\_count} are recorded as they existed at the specific retrieval timestamp, preserving the engagement state at the point of ingestion rather than maintaining a live synchronization with the source platform. 

\vspace{1em}

\noindent \textbf{\texttt{Actions}} are introduced to model dynamic interactivity, which functions as a relational record of the edges between \texttt{Accounts}, \texttt{Posts}, and \texttt{Communities}. This schema captures three types of interactions: content engagement (e.g., UPVOTE, SHARE, QUOTE), social ties (e.g., FOLLOW, BLOCK), and structural interlinks (LINK). This granular classification allows us to filter by \texttt{action\_type} to construct specific network topologies for downstream analysis. For instance, filtering for SHARE or QUOTE interactions facilitates the construction of information diffusion networks (similar to retweet graphs), which are instrumental in quantifying political polarization.

A critical architectural detail is the temporal dependency of the metadata: attributes like \texttt{like\_count} are recorded as they existed at the specific retrieval timestamp, preserving the engagement state at the point of ingestion rather than maintaining a live synchronization with the source platform. While logging both aggregate metrics in the \texttt{Posts} table and discrete interaction events in the \texttt{Actions} table may appear redundant from a strict database normalization perspective, this dual structure is necessary due to the inherent limitations of social media data collection. APIs and web scraping frameworks frequently expose macroscopic engagement metrics (e.g., total shares or likes) without providing access to the complete set of underlying microscopic network edges. Consequently, the quantitative metrics in the \texttt{Posts} table serve as an immutable macroscopic snapshot of engagement. In parallel, the \texttt{Actions} table logs the specific, granular actions and entities of \texttt{Actions} (such as a distinct SHARE or QUOTE between two identified users) strictly when that explicit relational data is available. This design ensures researchers can analyze total engagement volume even when the complete user-user interaction network cannot be fully reconstructed.

\vspace{1em}

\noindent \textbf{\texttt{Entities}} table functions as a granular repository for semantic and structural artifacts extracted from unstructured text fields, such as post bodies, community titles, and user biographies. By systematically indexing auxiliary components—including hashtags, URLs, user mentions, video/image media keys, and email addresses—this table bridges the gap between raw textual content and quantitative analysis. This structure is particularly critical for downstream tasks, such as tracking topic evolution via hashtags, mapping external information flows through URLs.

\vspace{1em}

\noindent \textbf{\texttt{AccountEnrichments}} stores supplemental metadata related to user profiles that cannot be directly mapped to the primary account table during standardization. Additionally, it serves as the storage layer for output generated by the various enrichment modules detailed in Section~\ref{section:enrichments_modules}. 

\vspace{1em}

\noindent \textbf{\texttt{PostEnrichments}} is similar to the \texttt{AccountEnrichments} table. This entity contains post-level metadata that falls outside the standard \texttt{Posts} table. It also captures and stores enriched data attributes produced by the post-processing modules.

\vspace{1em}

We implemented each of the overmentioned table schemas using PostgreSQL backend in a tabular format. The presence of the \texttt{created\_at} for all the tables is mandatory. Since, there can be duplicate records in the collected data, we basically capture the \texttt{retrieved\_at} of the records to enable the temporal changes of the records. The location entities in \texttt{Accounts} and \texttt{Posts} tables are geo coordinates and these coordinates can be obtained by using \texttt{geopy} library \citep{geopy} or any other alternatives. 

For each platform dataset, a dedicated adapter performs the required mapping and passes the generated actions and entities of \texttt{Actions} to our database helper. This helper manages bulk insertion and performs on-the-fly deduplication to systematically minimize database collisions. Researchers do not need to manage these underlying operations; they solely need to ensure their data mappings are accurately defined. Listing~\ref{lst:standardizer} illustrates the usage of the base \texttt{Standardizer} class. To utilize this class, the only required implementation is the \texttt{standardize} method, which accepts an \texttt{input\_record} argument. This argument contains both the raw record and its associated metadata from the source. For example, a record might be a Python dictionary representing a single Twitter data point, while the metadata contains contextual information such as the file path or line number. We have already developed several standardizer adapters for diverse platforms, fully referencing their respective datasets. Furthermore, by extending the base standardizer, researchers can implement custom adapters to utilize \projectname{} without managing complex database ingestion technicalities. All of these information are mentioned in the documentation of the software.

\begin{listing}[!ht] 
    \begin{mdframed}[style=fancyframe, frametitle={Standardizer Mapping}]
    \begin{lstlisting}[style=mystyle, language=Python] 
from typing import Any, List, Tuple
from smdt.standardizers.base import Standardizer, SourceInfo
from smdt.store.models import (Accounts, Posts, Actions, Communities, Entities, PostEnrichments, AccountEnrichments)

class MyCustomStandardizer(Standardizer):
    name = "my_custom_standardizer"

    def standardize(
        self, input_record: Tuple[dict, SourceInfo]) -> List[Any]:
        record, source_info = input_record
        
        output_models = []
        # Users will generate objects of the models and append the models to the output_models list
        return output_models
\end{lstlisting}
    \end{mdframed}
    \caption{A Sample usecase of the Standardizer Base Class}
    \label{lst:standardizer}
\end{listing}

\subsection{Inspector And Dataset Comparison}
\label{section:inspector}
To evaluate our data standardization methodology, we compiled several publicly available social media datasets. First, we examined the primary data entities and their representations across these collections. Table~\ref{tbl:platforms} presents the available standardized fields for each platform. A clear structural distinction emerges between threaded discussion platforms, such as Reddit, and micro blogging platforms like Twitter (X). As discussed for ``Communities,'' platforms like Reddit and Voat provide granular metadata for sub-communities (e.g., \textit{bio}, \textit{member\_count}). In contrast, platforms such as Twitter and Truth Social lack this community layer entirely, organizing content primarily around individual \textit{account\_ids}.

Furthermore, because some of these datasets were collected via web scraping rather than through official APIs, the absence of certain fundamental data types can be attributed to the methodological limitations inherent to scraping efforts. While the core text content and essential account details are generally obtainable across all platforms, specific temporal metadata is occasionally absent; for instance, datasets from Bluesky, Gab, and Scored lack the \textit{created\_at} timestamp for user accounts.

The \texttt{Actions} entity remains essential for capturing the diverse array of user interactions, such as sharing, quoting, or replying, which vary significantly in their technical implementation and data availability across these distinct platforms. We do not provide a direct comparison for this entity in the table. Because we aimed to avoid restricting our structure to the limitations of the currently available datasets, we designed the actions schema to comprehensively support the major types of interactions a user can perform on a \texttt{Community}, \texttt{Account}, or \texttt{Post}.

The \projectname{} framework operates strictly on an availability basis. It does not impute or synthetically generate missing metadata, such as absent \textit{created\_at} timestamps in certain scraped datasets. Downstream modules are designed to process available fields, and we rely on the Inspector component to help researchers audit variable availability before designing their cross-platform methodologies.

Conducting cross-platform analyses requires researchers to assess the availability of specific entities within standardized datasets. Previously, we compared this availability across datasets from previously mentioned platforms. However, that reference table may not be applicable when introducing new datasets or alternative platforms. 
To address this limitation, we introduce an \texttt{Inspector} component. Given a database connection (as each dataset is stored in a separate database), this module dynamically calculates the statistics for various entities within each standardized table. Listing~\ref{lst:inspector} provides a code example demonstrating the usage of this module. By instantiating \texttt{Inspector} objects for two distinct databases, users can invoke the \texttt{report\_schemas} function to generate a side-by-side, pretty-printed statistical summary of each table. This functionality allows researchers to quickly observe the available data and determine the types of high-level analyses that are feasible across the selected datasets.

\begin{listing}[H] 
    \begin{mdframed}[style=fancyframe, frametitle={Inspector}]
    \begin{lstlisting}[style=mystyle, language=Python] 
from smdt.store.standard_db import StandardDB
from smdt.inspector import Inspector, report_schemas

# 1. Connect to your databases
db_name1 = "my_local_db_1"
db_name2 = "my_local_db_2"

sdb1 = StandardDB(db_name1, initialize=False)
sdb2 = StandardDB(db_name2, initialize=False)

# 2. Create an Inspector instance
# You can inspect the 'public' schema or any other schema you use
inspector1 = Inspector(sdb1, schema="public")
inspector2 = Inspector(sdb2, schema="public")

# 3. Generate and print the report
# The report_schemas function handles formatting and printing to stdout
# Pass a list of inspectors to compare multiple DBs or schemas if needed
report_schemas([inspector1, inspector2], only_tables=["posts", "accounts"])
\end{lstlisting}
    \end{mdframed}
    \caption{Example usage of the Inspector module to generate a comparative statistical report across two distinct datasets.} 
    \label{lst:inspector}
\end{listing}

\begin{table}[h!]
    \centering
    \begin{threeparttable}
    \rowcolors{2}{white}{rowgray}
    
    \caption{\textbf{Comparison of Platform Data Fields.} We present data fields available for different platforms and public datasets.}
    \label{tbl:platforms}
    
    \renewcommand{\arraystretch}{1} 
    
    \fontsize{9pt}{9pt}\selectfont
    \begin{tabular}{l l c c c c c c c c c c c c c}
        \toprule
        \rowcolor{white} 
        & & \rotatebox[origin=l]{90}{Twitter V1             \citep{najafi2024first}} & 
         \rotatebox[origin=l]{90}{Twitter V2 \citep{najafi2024first}} & 
         \rotatebox[origin=l]{90}{BlueSky \citep{failla2024m}} &
         \rotatebox[origin=l]{90}{Gab \citep{gab_posts_2018}} &
         \rotatebox[origin=l]{90}{VoatCo \citep{mekacher2022can}} &
         \rotatebox[origin=l]{90}{Scored \citep{patel2024idrama}} &
         \rotatebox[origin=l]{90}{TruthSocial \citep{gerard2023truth}} &
         \rotatebox[origin=l]{90}{Koo \citep{mekacher2024koo}} &
         \rotatebox[origin=l]{90}{Reddit \citep{baumgartner2020pushshift}} &
         \rotatebox[origin=l]{90}{USC-TruthSocial \citep{shah2024unfiltered}} &
         \rotatebox[origin=l]{90}{USC-X \citep{balasubramanian2024public}} &
         \rotatebox[origin=l]{90}{Telegram \citep{baumgartner2020pushshift}} & 
         \rotatebox[origin=l]{90}{Parler \citep{aliapoulios2021early}} 
          \\
        \midrule

        \multicolumn{2}{l}{\textbf{Communities}} & & & & & & & & & & & & & \\
          & \texttt{community\_id} & \no & \no & \no & \no & \yes & \yes & \no & \no & \yes & \no & \no  & \yes & \no \\ 
          & \texttt{community\_username} & \no & \no & \no & \no & \yes & \yes & \no & \no & \yes & \no & \no  & \yes & \no \\ 
          & \texttt{community\_name} & \no & \no & \no & \no & \yes & \no & \no & \no & \yes & \no & \no  & \yes & \no \\ 
          & \texttt{bio} & \no & \no & \no & \no & \yes & \yes & \no & \no & \yes & \no & \no  & \yes & \no \\ 
          & \texttt{is\_public} & \no & \no & \no & \no & \yes & \no & \no & \no & \yes & \no & \no  & \yes & \no \\ 
          & \texttt{member\_count}& \no & \no & \no & \no & \yes & \no & \no & \no & \yes & \no & \no  & \yes & \no \\ 
          & \texttt{post\_count} & \no & \no & \no & \no & \no & \no & \no & \no & \no & \no & \no  & \yes & \no \\ 
          & \texttt{profile\_image\_url} & \no & \no & \no & \no & \no & \no & \no & \no & \yes & \no & \no  & \no & \no \\ 
          & \texttt{owner\_account\_id} & \no & \no & \no & \no & \yes & \yes & \no & \no & \no & \no & \no  & \no & \no \\ 
          & \texttt{created\_at} & \no & \no & \no & \no & \yes & \yes & \no & \no & \yes & \no & \no  & \yes & \no \\ 
          & \texttt{retrieved\_at} & \no & \no & \no & \no & \yes & \no & \no & \no & \yes & \no & \no  & \yes &  \no\\ 

        \midrule
         
        \multicolumn{2}{l}{\textbf{Accounts}} & & & & & & & & & & & & & \\
        & \texttt{account\_id}      & \yes & \yes & \yes & \yes & \yes & \yes & \yes & \yes & \yes & \yes & \yes  & \yes & \yes \\ 
        & \texttt{user\_name}       & \yes & \yes & \no  & \yes & \no & \no  & \yes & \yes & \yes & \yes & \yes  & \yes & \yes \\ 
        & \texttt{profile\_name}    & \yes & \yes & \no  & \yes & \no  & \no  & \no  & \no  & \yes & \no & \yes  & \yes & \no \\ 
        & \texttt{bio}              & \yes  & \yes  & \no  & \no  & \no  & \no  & \no  & \yes  & \no  & \no & \yes   & \no & \yes \\ 
        & \texttt{location}         & \no  & \no  & \no  & \no  & \no  & \no  & \no  & \no  & \no  & \no & \no  & \no & \no \\ 
        & \texttt{post\_count}      & \yes & \yes & \no & \yes & \no & \no & \yes & \no & \no & \no & \yes  & \no & \yes \\ 
        & \texttt{friend\_count}    & \yes & \yes & \no  & \no  & \no  & \no  & \yes & \no  & \no  & \no & \yes  & \no & \yes \\ 
        & \texttt{follower\_count}  & \yes & \yes & \no  & \no  & \no  & \no  & \yes & \no  & \no  & \no & \yes  & \no & \yes \\ 
        & \texttt{is\_verified}  & \yes & \yes & \no  & \yes  & \no  & \no  & \no & \no  & \no  & \no & \yes  & \yes & \yes \\ 
        & \texttt{profile\_image\_url}& \yes  & \yes  & \no  & \yes  & \no  & \no  & \no  & \no  & \no  & \no & \yes & \no  &  \no\\ 
        & \texttt{created\_at}      & \yes & \yes & \no  & \no  & \yes & \no  & \yes & \yes & \yes & \yes & \yes & \yes & \yes \\ 
        & \texttt{retrieved\_at}    & \no  & \yes & \no  & \yes & \no  & \no  & \yes & \no  & \yes & \yes & \no & \yes & \no \\ 
        \midrule
        
        \multicolumn{2}{l}{\textbf{Posts}} & & & & & & & & & & & & & \\
        & \texttt{account\_id}      & \yes & \yes & \yes & \yes & \yes & \yes & \yes & \yes & \yes & \yes & \yes  & \yes & \yes \\ 
        & \texttt{post\_id}         & \yes & \yes & \yes & \yes & \yes & \yes & \yes & \yes & \yes & \yes & \yes  & \yes & \yes \\ 
        & \texttt{conversation\_id} & \no  & \yes  & \no  & \yes  & \no  & \no  & \no  & \no  & \yes  & \no & \yes  & \no & \no \\ 
        & \texttt{community\_id} & \no  & \no  & \no  & \no  & \yes  & \yes  & \no  & \no  & \yes  & \no & \no  & \yes & \no \\ 
        & \texttt{body}             & \yes & \yes & \yes & \yes & \yes & \yes & \yes & \yes & \yes & \yes & \yes  & \yes & \yes \\ 
        & \texttt{location}         & \yes  & \yes  & \no  & \no  & \no  & \no  & \no  & \no  & \no  & \no & \no  & \no & \no \\ 
        & \texttt{like\_count}      & \yes  & \yes  & \no  & \yes  & \no  & \yes  & \yes  & \no  & \yes  & \yes & \yes  & \no & \yes \\ 
        & \texttt{dislike\_count}      & \no  & \no  & \no  & \yes  & \no  & \yes  & \no  & \no  & \no  & \no & \no & \no & \no \\ 
        & \texttt{view\_count}      & \no  & \yes  & \no  & \no  & \no  & \no  & \no  & \no  & \no  & \yes & \yes  & \no & \yes \\ 
        & \texttt{share\_count}     & \yes  & \yes  & \yes  & \yes  & \no  & \no  & \yes  & \no  & \no  & \yes & \yes & \no & \yes \\ 
        & \texttt{comment\_count}   & \yes  & \yes  & \yes  & \yes  & \no  & \no  & \yes  & \no  & \no  & \yes & \yes  & \no & \yes \\ 
        & \texttt{quote\_count}     & \yes  & \yes  & \no  & \no  & \no  & \no  & \no  & \no  & \no  & \no & \yes  & \no & \no \\ 
        & \texttt{bookmark\_count}  & \no  & \no  & \no  & \no  & \no  & \no  & \no  & \no  & \no  & \no & \no  & \no & \no \\ 
        & \texttt{created\_at}      & \yes & \yes & \yes & \yes & \yes & \yes & \yes & \yes & \yes & \yes & \yes & \yes & \yes \\ 
        & \texttt{retrieved\_at}    & \no  & \yes & \no  & \yes & \no  & \no  & \yes & \no  & \yes & \yes & \no  & \yes & \no \\ 



        \bottomrule
    \end{tabular}
    \end{threeparttable}
\end{table}

\subsection{Data Pseudonymization}
While standardized datasets provide critical insights, they often contain sensitive Personally Identifiable Information (PII)—including usernames, email addresses, and unique identifiers—that poses a risk of user re-identification. To address this, we implemented a comprehensive data anonymization module designed to secure the dataset by cryptographically hashing PII. Leveraging the \texttt{whirlpool}\footnote{\href{https://github.com/oohlaf/python-whirlpool}{https://github.com/oohlaf/python-whirlpool}} algorithm alongside Python's \texttt{hashlib} library\footnote{\url{https://docs.python.org/3/library/hashlib.html}}, the module supports a diverse array of secure hash and message digest algorithms. The system prioritizes configurability, enabling users to specify both the hashing algorithm and a custom cryptographic salt. This protocol is applied to all database identifiers (IDs) to ensure consistent relational mapping, while simultaneously redacting sensitive entities detected within post bodies and other unstructured textual fields. Listing~\ref{lst:anonymization} provides a practical example of applying these cryptographic hashing and redaction techniques.

\begin{listing}[!ht] 
    \begin{mdframed}[style=fancyframe, frametitle={Sample Anonymization Code Snippet}]
    \begin{lstlisting}[style=mystyle, language=Python] 
from smdt.config import AnonymizationVariables
from smdt.anonymizer import Anonymizer, AnonymizeConfig, Algorithm, DEFAULT_POLICY

anon_vars = AnonymizationVariables()
cfg = AnonymizeConfig(
    src_db_name="my_local_db", dst_db_name="my_local_db_anon", pepper=anon_vars.pepper, 
    algorithm=Algorithm.SHA256, output_hex_len=64, ask_reinit=True, chunk_rows=5_000
)
az = Anonymizer(cfg, DEFAULT_POLICY)
az.run()
\end{lstlisting}
    \end{mdframed}
    \caption{\textbf{Anonymization} A sample code for anonymizing the dataset with SHA256 algorithm with the provided pepper.}
    \label{lst:anonymization}
\end{listing}

\FloatBarrier
\subsection{Enrichment Modules}
\label{section:enrichments_modules}

Raw social media data often lacks the depth required for comprehensive research, making it necessary to enrich the foundational data with additional context about accounts and content. For example, complex tasks such as detecting coordinated activities and tracking online campaigns rely heavily on this supplementary information. To support this, our framework incorporates external tools and machine learning models designed for specific tasks, treating the information obtained from these additional models as an enrichment of the collected data.

Examples of data enrichment includes BotometerLite for social bot detection \citep{yang2020scalable} and M3Inference model for estimating user demographics \citep{wang2019demographic}. One can consider additional models to infer account-level details like inferring location from profile descriptions, fake personas from profile pictures, or detecting anomalous followers from social network topology \citep{zouzou2024unsupervised}.

Social media content also offers valuable information and models for detecting sentiment, hate speech or toxicity can be applied to textual information \citep{tweetnlp,Detoxify}. More sophisticated models can incorporate links shared with the tweets to estimate information credibility. Images or videos attached to the content can be also processed with multi-model approaches of fake news detection.

All these account- and content-level enrichment can be used to study information cascades and coordination among accounts. We offer few analysis modules that process base information and enrichment; but we want to emphasize \projectname{} is highly flexible to extend with other existing models and users' own methodologies. 

\subsubsection{Social Networks Module}

The Social Networks Module is designed to construct tailored networks for analyzing information diffusion. It enables users to generate three primary network architectures:

\begin{itemize}
    \item \textbf{User–User Interaction Networks:} Mapping direct interactions between individuals (e.g., SHARE and QUOTE networks). 
    
    \item \textbf{Co-occurrence Networks:} Identifying the shared usage of specific features, such as hashtags and mentions.
    
    \item \textbf{Bipartite Networks:} Modeling relationships between two distinct entity types (e.g., accounts and shared hashtags or domains). 

\end{itemize}

Additionally, the module supports temporal network construction, allowing users to analyze structural evolution over time and track the dynamic spread of information. Listing~\ref{lst:share_net_builder} demonstrates how to build temporal share networks using a standardized dataset.

\begin{listing}[!ht] 
    \begin{mdframed}[style=fancyframe, frametitle={Share network generation}]
    \begin{lstlisting}[style=mystyle, language=Python] 
from datetime import datetime, timedelta
from smdt.store.standard_db import StandardDB
from smdt import networks

db = StandardDB("my_local_db")

windows = networks.user_interaction_over_time(
    db,
    interaction="SHARE",
    start_time=datetime(2023, 5, 14),
    end_time=datetime(2023, 5, 15),
    step=timedelta(hours=1),
    weighting="count",
    min_weight=3,
)

for win in windows:
    ws = win["window_start"]
    we = win["window_end"]
    net = win["network"]
    print(ws, we, net.meta["edge_count"])
\end{lstlisting}
    \end{mdframed}
    \caption{Temporal SHARE (Retweet) Network Generation}
    \label{lst:share_net_builder}
\end{listing}

\subsubsection{Natural Language Processing Module}

We provide an advanced enrichment layer for standardized datasets, allowing users to apply both online and offline NLP models directly to their data. The offline sublayer enables the local execution of Hugging Face models using standard model identifiers, whereas the online sublayer provides a unified interface for external providers such as Gemini, Claude, and OpenAI, alongside local vLLM servers. This flexible architecture empowers users to seamlessly deploy arbitrary models—ranging from sentiment and hate speech classifiers to custom, prompt-specific generative tasks. Listing \ref{lst:nlp} demonstrates the configuration required to run an Anthropic Claude model through this enrichment pipeline.

\begin{listing}[!t] 
    \begin{mdframed}[style=fancyframe, frametitle={TextGen enricher}]
    \begin{lstlisting}[style=mystyle, language=Python] 
from smdt.store.standard_db import StandardDB
from smdt.enrichers.runner import run_enricher

db_name = "my_local_db"
db = StandardDB(db_name=db_name)

api_key = "API_KEY"

config = {
    "model_id_postfix": "v1_sentiment_claude",   
    "chat_model_id": "claude-3-5-sonnet-20241022", 
    "base_url": "https://api.anthropic.com/v1/messages",
    "api_key": api_key,
    "provider_kind": "anthropic",
    # The Prompt
    "system_prompt": "You are a helpful assistant.",
    "user_template": "Analyze the sentiment of this post: {body}",
    # Settings
    "only_missing": True,  
    "batch_size": 10,
    "reset_cache": True,
    "max_tokens": 1000,
}

run_enricher("textgen", db=db, **config)
\end{lstlisting}
    \end{mdframed}
    \caption{TextGen Enricher configuration example utilizing Anthropic's Claude model.}
    \label{lst:nlp}
\end{listing}





\subsubsection{Bot Detection}

Behavioral anomalies and malicious intentions are common for all platform. However, their strategies and mechanisms may change. In literature, there are models trained on labeled datasets for particular platforms like Twitter. Botometer is one of the well-known model for Twitter \citep{varol2017online}, but this model extracts platform specific features to capture diverse automated behavior. Because of that this model will be only usable for Twitter specific datasets upon the availability of the profile level information.

\section{Case studies}

In this section, we demonstrate the versatility of \projectname{} through a series of selected case studies. Each subsequent subsection utilizes distinct configurations to showcase the specific functionalities and performance of our system. Datasets analyzed in these studies comes from different APIs, or format due to scraping; however, \projectname{} standardize these data and apply the same tools for downstream tasks.
 
\subsection{Comparative Analysis of Stance Towards the Russia-Ukraine War}

We selected the \textit{TwitterArchive} and \textit{TruthSocial} datasets due to their substantial political discourse regarding the Russia–Ukraine war. Following the standardization of the \textit{USC-TruthSocial} and \textit{USC-X} datasets, we filtered the posts using war-related keywords (listed in the Appendix). This process yielded approximately $58{,}000$ posts from TruthSocial and roughly $1.74$ million posts (thirty times the volume) from Twitter. To mitigate this significant volume disparity and ensure balanced comparisons, we constructed $30$ bootstrapped samples from the Twitter dataset, each containing $58{,}000$ posts. Next, we deployed the \texttt{deepseek-ai/deepseek-llm-7b-chat}\footnote{\href{https://huggingface.co/deepseek-ai/deepseek-llm-7b-chat}{https://huggingface.co/deepseek-ai/deepseek-llm-7b-chat}} model via the vLLM tool within our enrichment layer to identify each post's stance toward the conflict. The model was prompted to classify each post into one of five categories: Pro-Ukraine (expressing support for Ukraine, its government, military, or actions), Anti-Ukraine (criticizing or opposing Ukraine), Pro-Russia (supporting Russia, its government, military, or actions), Anti-Russia (criticizing or opposing Russia), or Neutral (showing no clear stance or not directly addressing the conflict). To validate this stance detection pipeline, a human annotator labeled $300$ posts from each platform ($600$ total), upon which the pipeline achieved an F1-score of $0.62$. Finally, to compare the ideological leanings across the two platforms, we conducted two-proportion Z-tests using the stance distributions from each of the $30$ sampled batches. In every iteration, the p-values fell below the conventional significance threshold of $0.05$. This indicates that TruthSocial consistently contained a significantly higher proportion of Russia-leaning posts than Twitter, a finding that holds true even after rigorously accounting for sampling variability.

\subsection{Temporal Sentiment Analysis}

We standardized four distinct social media datasets for our analysis: \textit{USC-TruthSocial}, \textit{USC-X}, the 2023 Turkish General Election dataset \citep{najafi2024first}, and a 2023 Finnish Parliamentary Election dataset. To ensure linguistic accuracy across these diverse sources, we applied language-specific sentiment models. For the English datasets, we utilized \texttt{bertweet-base-sentiment-analysis}\footnote{\href{https://huggingface.co/finiteautomata/bertweet-base-sentiment-analysis}{huggingface.co/finiteautomata/bertweet-base-sentiment-analysis}}, optimized for informal syntax and emojis. For the Turkish dataset, we applied \texttt{TurkishBERTweet-Lora-SA}\footnote{\href{https://huggingface.co/VRLLab/TurkishBERTweet-Lora-SA}{huggingface.co/VRLLab/TurkishBERTweet-Lora-SA}}, a LoRA-fine-tuned transformer tailored for Turkish discourse \citep{najafi2024turkishbertweet}. Finally, for the Finnish dataset, we employed \texttt{bert-finnish-sentiment-analysis}\footnote{\href{https://huggingface.co/nisancoskun/bert-finnish-sentiment-analysis}{huggingface.co/nisancoskun/bert-finnish-sentiment-analysis}}, fine-tuned for high precision in low-resource contexts.

\begin{figure}[t]
    \centering
    \includegraphics[width=\linewidth]{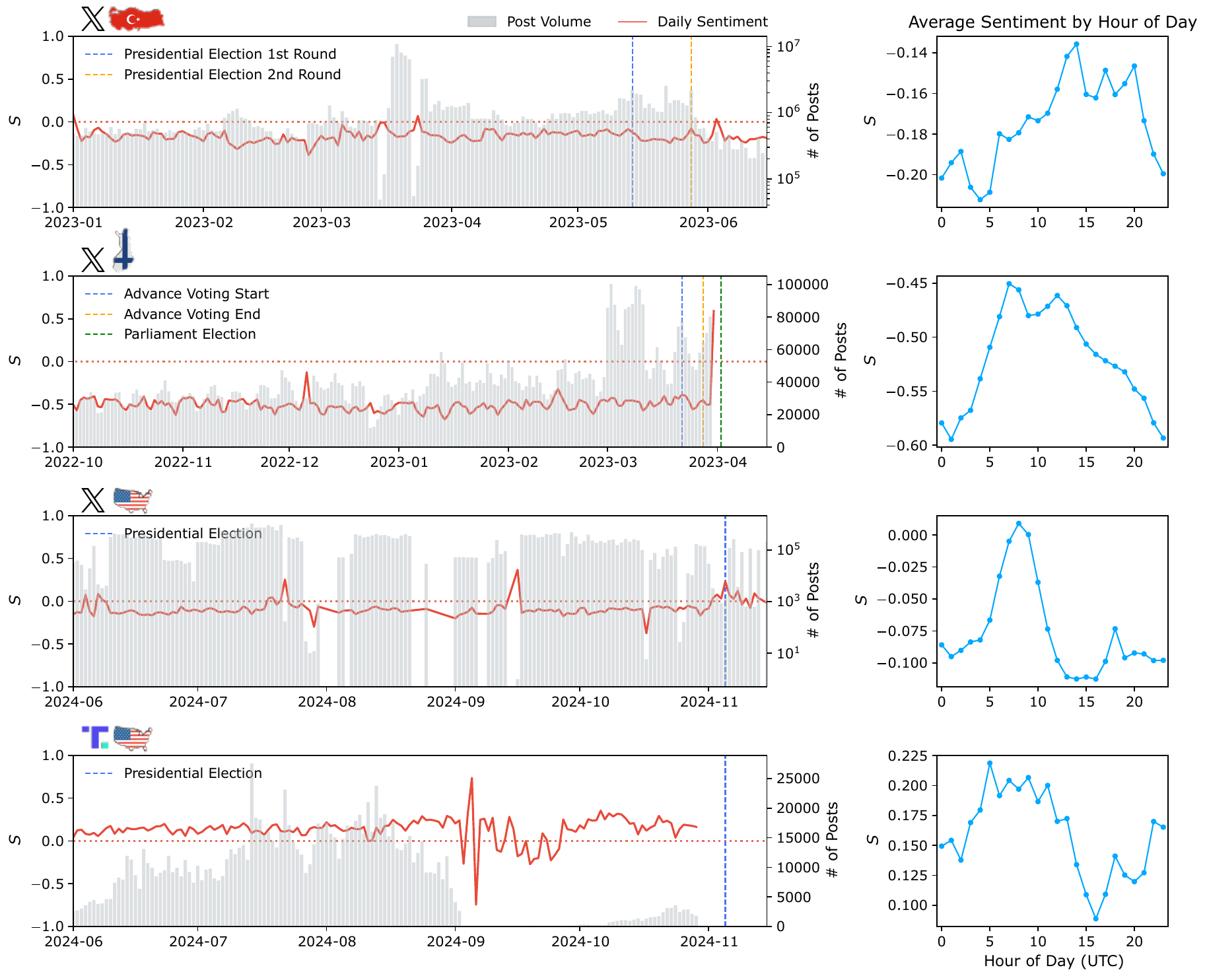}
    \caption{Daily sentiment and post volume surrounding key election events, with corresponding hourly sentiment trends. $S$ indicates the average sentiment of posts.}    
    \label{fig:sent_temporal}
\vspace{1em}
\end{figure}

Figure \ref{fig:sent_temporal} illustrates the temporal dynamics of daily sentiment and post volumes for each dataset, spanning from approximately four to six months prior to their respective elections. The left-hand panels display both aggregated daily sentiment and daily post volume. In the Turkish dataset, which begins in January 2023, generally we have aggregated negative sentiments for all of the days. Specially, there is a decline in the negative sentiment around February 2023 and increase in the number of posts; this reflects the impact of the regional earthquakes. Additionally, the drop observed in the March 2023 post volume data coincides with Twitter's structural migration from API v1 to API v2.

The Finnish dataset exhibits a distinct, cyclical posting frequency. This pattern shows a marked increase beginning in the first two weeks of March 2023, driven by the lead-up to advance voting. As a large portion of the Finnish electorate votes early, activity naturally peaked around the official advance voting period from March 22 to March 28. Meanwhile, the US datasets (\textit{USC-X} and \textit{USC-TruthSocial}) demonstrate more volatile posting patterns. This variance is an expected artifact of their compilation via web scraping, which introduces more noise compared to the API-based retrieval methods used for the Turkish and Finnish datasets. Furthermore, the \textit{USC-TruthSocial} dataset contains a lower total post volume than the others, further contributing to its structural variance.

The hourly plots reveal the intra-day variations of average sentiment for each dataset. When standardized to UTC, distinct patterns emerge reflecting geographic and platform-specific behaviors. Turkish Twitter sentiment peaks between approximately 10:00 and 15:00 UTC. Finnish Twitter, while maintaining a lower overall sentiment baseline, exhibits a peak during the morning hours. Comparing the two US platforms reveals diverging intra-day dynamics: \textit{USC-X} sentiment reaches its maximum around 05:00 UTC before decreasing , whereas \textit{USC-TruthSocial} sustains positive sentiment for a longer duration, featuring a distinct minimum at 16:00 UTC. These diurnal variations suggest that sentiment on social media corresponds closely to the daily behavioral cycles of its users.

\subsection{Cross-Platform Hashtag and Domain Analysis}
\begin{figure}[!t]
    \includegraphics[width=\linewidth]{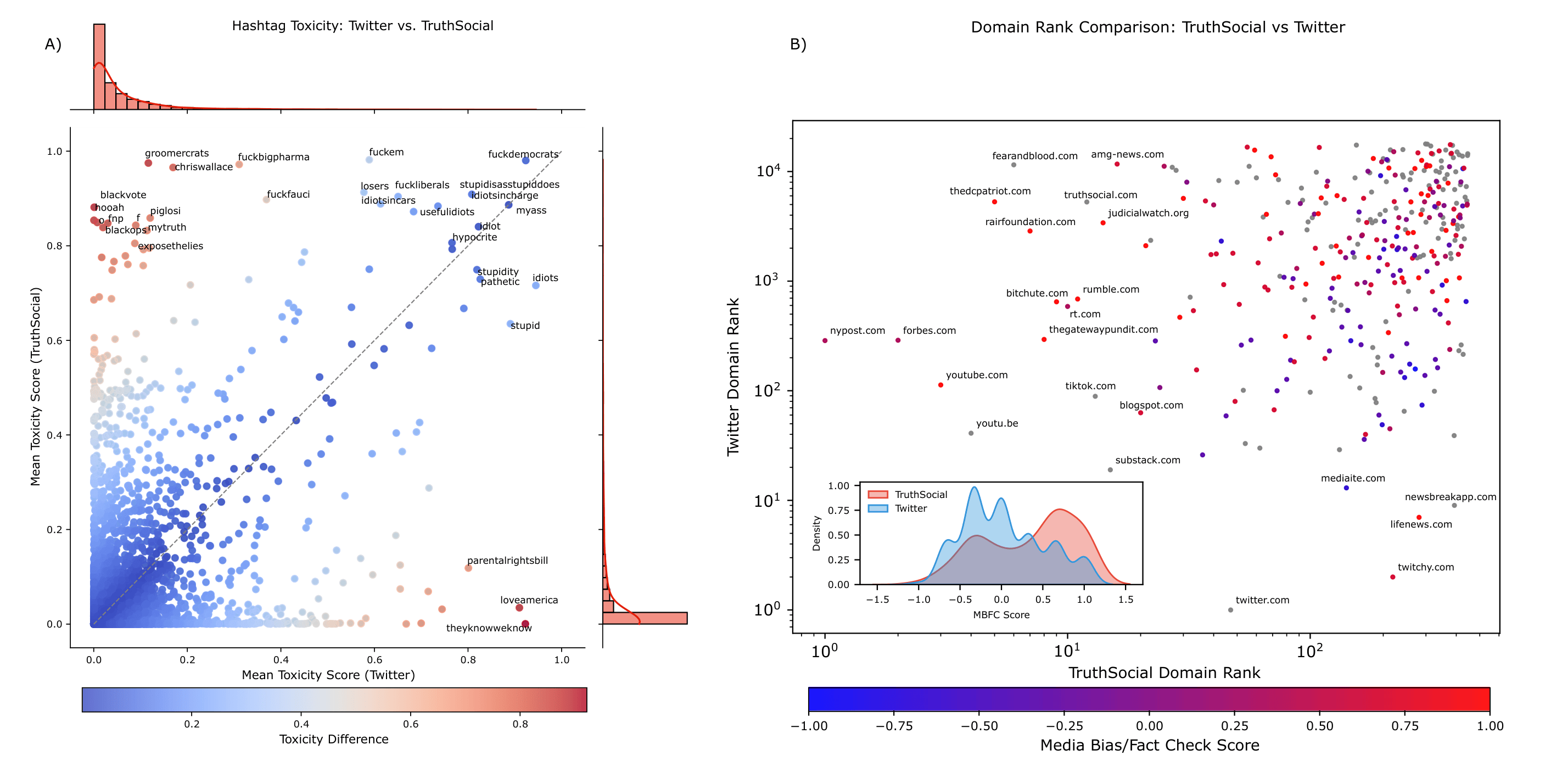}
    \caption{\textbf{Hashtag and Domain Analyses Between Twitter and TruthSocial.} 
    A) \textbf{Hashtag Toxicity Comparison.} Each point represents a hashtag appearing at least twice on both platforms. The x-axis shows the mean toxicity score on Twitter, and the y-axis shows the mean score on TruthSocial. Color indicates the absolute difference in toxicity between two platforms.
    B) \textbf{Domain Popularity and Bias.} Each point represents a domain, positioned by its popularity rank on TruthSocial (x-axis) and Twitter (y-axis). Lower ranks correspond to higher popularity. Color reflects the Media Bias/Fact Check (MBFC) score, from left-leaning (blue) to right-leaning (red). The inset shows the distribution of MBFC scores per platform, illustrating a rightward shift in domain bias on TruthSocial.}
    \label{fig:hashtag_toxicity_comparison}
\end{figure}

Following the standardization of the \textit{TruthSocial} \citep{gerard2023truth} and \textit{TwitterArchive} datasets, we extracted the hashtags and domains shared across both platforms. To calculate post-level toxicity scores, we utilized the Detoxify module \citep{Detoxify}—which provides pre-trained models for toxic comment prediction—and subsequently aggregated these metrics by hashtag. The left panel of Figure \ref{fig:hashtag_toxicity_comparison} illustrates the temporal alignment between the two platforms regarding hashtag usage and their respective toxicity differentials. Furthermore, the right panel of Figure \ref{fig:hashtag_toxicity_comparison} presents the domain rankings alongside their corresponding Media Bias/Fact Check (MBFC) scores. These MBFC scores were sourced from prior work \citep{yang2025domaindemo} and are publicly available via GitHub\footnote{\href{https://github.com/LazerLab/DomainDemo/blob/main/data/existing\_labels/partisan\_bias\_scores.csv}{https://github.com/LazerLab/DomainDemo/blob/main/data/existing\_labels/partisan\_bias\_scores.csv}}.

The distribution of hashtag toxicity across both platforms reveals a significant concentration of low-toxicity content, as evidenced by the dense clustering of data points near the origin $(0,0)$ and the strongly right-skewed marginal density distributions on the top and right axes. While the majority of hashtags remain relatively benign, the color gradient representing toxicity differences highlights notable platform-specific disparities. Specifically, a pronounced cluster in the upper-left region identifies hashtags that are highly toxic on TruthSocial (scores $>0.8$) but yield very low toxicity on Twitter (scores $<0.2$). These are frequently politically charged or targeted at specific individuals, such as \textit{\#groomercrats} and \textit{\#piglosi}. Conversely, a smaller subset of hashtags in the lower-right region demonstrates high toxicity on Twitter while remaining low on TruthSocial. Hashtags characterized by general profanity or broad insults tend to fall closer to the upper-right diagonal, indicating a shared level of high toxicity across both ecosystems.

This divergence in toxicity is accompanied by a stark ideological divide in shared media, characterized by polarized domain preferences across the platforms. As illustrated by the inset density plot, Twitter’s domain distribution is bimodal, centering primarily within the negative-to-neutral range of the MBFC scores, which indicates a prevalence of left-leaning to moderate sources. In contrast, the distribution on TruthSocial is heavily concentrated in the positive range, signaling a dominant preference for right-leaning or conservative domains. This asymmetrical popularity is further evidenced in the scatter plot by a dense concentration of high-positive MBFC scores in the upper-left quadrant. High-ranking domains on TruthSocial, such as \textit{thegatewaypundit.com}, \textit{bitchute.com}, and \textit{rt.com}, exhibit significantly lower visibility on Twitter. While mainstream outlets and traditional news platforms like \textit{youtube.com} and \textit{forbes.com} maintain high ranks across both ecosystems, left-leaning or centrist domains generally cluster toward the lower-right, demonstrating consistently stronger performance on Twitter than on TruthSocial.

\subsection{Temporal Hashtag Co-Occurrence Analyses}
\begin{figure}[!t]
    \centering
        \includegraphics[width=\linewidth]{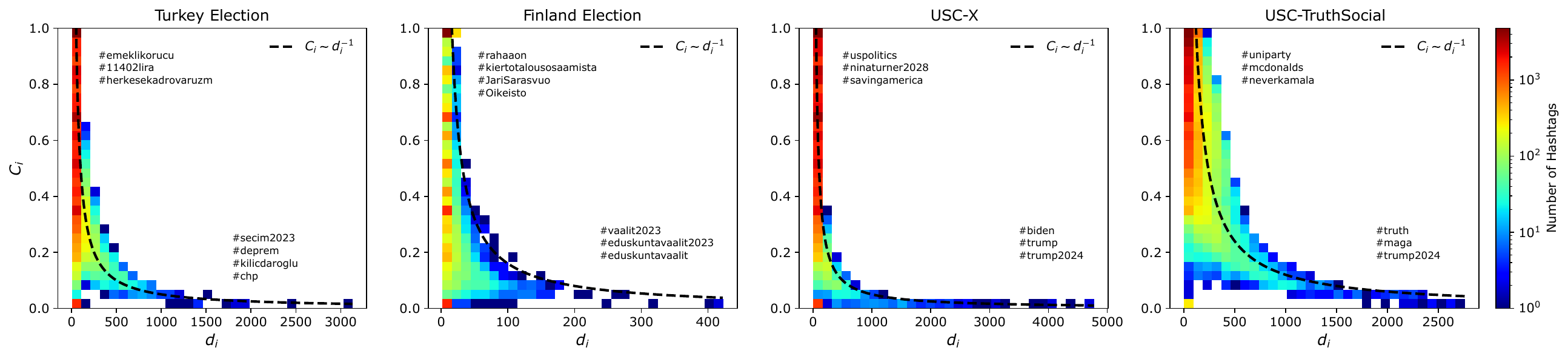}
     \caption{\textbf{Relationship between local clustering coefficient and node degree.} Analysis conducted to investigate four different datasets. High-level hashtags for each context is identified and highlighted in figures.}
     \label{fig:generic_vs_specific_hashtags}
\end{figure}

We replicated the analysis of node clustering centrality versus degree as described by \citet{lorenz2018tracking}, applying it to four datasets: Finnish and Turkish elections, \textit{USC-X}, and \textit{USC-TruthSocial}. To capture the temporal dynamics of the discourse, we constructed weekly hashtag co-occurrence networks, performed modularity-based community detection, and computed clustering centrality metrics. The results are illustrated in Figure \ref{fig:generic_vs_specific_hashtags}. As expected in hierarchical networks, low clustering coefficients characterize top-level hashtags associated with broad topics, while the upper part of the distribution comprises highly specific, niche hashtags.

In the Turkey Election dataset, high-level hashtags include election-centric terms such as \texttt{\#secim2023}, \texttt{\#kilicdaroglu}, and \texttt{\#chp}, alongside \texttt{\#deprem} (earthquake), which reflects the profound impact of the 2023 Earthquakes in south-east Turkey on the broader public discourse. Conversely, specific hashtags with high clustering in this dataset, such as \texttt{\#emeklikorucu}, \texttt{\#herkesekadrovaruzm}, and \texttt{\#11402lira}, correlate with highly focused discussions regarding social security and economic issues. For the Finland Election dataset, broad, high-level hashtags are composed of general electoral terms like \texttt{\#vaalit2023}, \texttt{\#eduskuntavaalit2023}, and \texttt{\#eduskuntavaalit}. In contrast, highly clustered, specific hashtags in this dataset highlight targeted topics and figures, including \texttt{\#rahaaon}, \texttt{\#kiertotalousosaamista}, \texttt{\#JariSarasvuo}, and \texttt{\#Oikeisto}. Similarly, the \textit{USC-X} and \textit{USC-TruthSocial} datasets feature broad, high-level hashtags related to the US elections, including \texttt{\#trump}, \texttt{\#biden}, and \texttt{\#maga}. In contrast, the highly specific hashtags within these US datasets consist of targeted campaign or political markers, such as \texttt{\#uspolitics}, \texttt{\#ninaturner2028}, and \texttt{\#savingamerica}.

These analysis demonstrates \projectname{} capabilities to parallelize analysis pipelines across multiple datasets without any additional effort and how it scales with large-datasets. This framework offers different modules to study content, accounts, and networks.

\section{Discussion and Future Work}

To enable robust cross-platform comparisons, our framework maps diverse entity types to a uniform specification. This creates a shared analytical foundation that eliminates the structural and linguistic discrepancies often hindering collaborative research and data integration. By adhering to this standardized framework, researchers can ensure methodological consistency in their final evaluations, mitigating the risk of analytical errors stemming from data heterogeneity. Furthermore, our modular approach, which incorporates various user-level and content-level enrichment tools, enables researchers from diverse disciplinary backgrounds to seamlessly integrate these analyses into their workflows.

The necessity of such standardized analytical tools is underscored by recent regulatory shifts. Since 2018, the European Union has moved to regulate social media environments through frameworks such as the Digital Services Act (DSA). These regulations aim to combat the misuse of digital platforms, specifically addressing the drivers of online polarization, the dissemination of state-sponsored propaganda, and the systemic threats to global democratic processes. By providing a transparent and unified framework for data analysis, our work supports the overarching goals of these legislative efforts, offering a technical means to monitor and evaluate the digital ecosystem with greater precision.

While the current platform effectively supports the storage of statistical data, textual content, and file-based references for visual media, there are several avenues for future enhancement. A primary improvement involves integrating vector databases, such as ChromaDB \citep{chromadb}, to support semantic enrichment and enable complex semantic queries. Although we currently store semantic features within enrichment tables in our standardized database, transitioning to a dedicated vector architecture would provide significantly greater flexibility and retrieval efficiency. 

Additionally, driven by recent advancements in artificial intelligence and the rise of agentic models standardized under the Model Context Protocol (MCP) \citep{mcp_intro}, our framework is well-positioned to serve as a critical bridge. By connecting distinct layers of social media data, it can reliably support the creation of advanced analytical agents capable of navigating and interpreting complex multi-platform environments.

\textit{Limitations} While this work introduces a standardized infrastructure that enables researchers to conduct cross-platform analyses and allows non-technical users to leverage advanced enrichment tools, the current software is an early-stage Online Analytical Processing (OLAP) system. Future optimization is required to enhance its scalability and multi-dimensional flexibility. Additionally, the scope of cross-platform analysis is inherently constrained by external data availability. Consequently, we restricted our evaluation to the aforementioned case studies to demonstrate the system's baseline usability.

\section*{Acknowledgments}
The authors would like to acknowledge the support of CHIST-ERA for the CON-NET project and their local funding agencies for this grant.
The agencies are the Scientific and Technological Research Council of Türkiye (grant: 222N311), the Research Council of Finland (grant: 357743).

\bibliographystyle{unsrt}  
\bibliography{references}

\end{document}